# Near-ionization-threshold emission in atomic gases driven by intense sub-cycle pulses

Wei-Chun Chu, John C. Travers, and Philip St.J. Russell

*Max Planck Institute for the Science of Light, Günther-Scharowsky-Straße 1, D-91058 Erlangen, Germany*

wei-chun.chu@mpl.mpg.de

Abstract. We study theoretically the dipole radiation of a hydrogen atom driven by an intense sub-cycle pulse. The time-dependent Schrödinger equation for the system is solved by *ab initio* calculation to obtain the dipole response. Remarkably, a narrowband emission lasting longer than the driving pulse appears at a frequency just above the ionization threshold. An additional calculation using the strong field approximation also recovers this emission, which suggests that it corresponds to the oscillation of nearly-bound electrons that behave similarly to Rydberg electrons. The predicted phenomenon is unique to ultrashort driving pulses but not specific to any particular atomic structure.

PACS: 42.65.Ky, 42.65.Re, 32.80.Rm

## 1. Introduction

The electronic responses of atomic or molecular systems to laser fields in the strong-field regime (i.e., peak intensities higher than $10^{14}$ W/cm$^2$ in the near infrared) have been intensively studied over the past two decades [1,2]. In particular, the radiation from such a response is commonly studied in the context of high-order harmonic generation (HHG), where bursts of extreme ultraviolet light occur at every half optical cycle along the driving field, forming an attosecond pulse train. In recent years, the production of isolated attosecond pulses (as opposed to pulse trains) through HHG has been of great interest for research in ultrafast phenomena, and has created the new field of attoscience [2]. As truly single-cycle or sub-cycle driving pulses are not readily available, attosecond pulses are usually generated using a variety of gating schemes [3], but are not in widespread use due to high experimental complexity. Instead of using these gating techniques, one could decrease the length of the driving pulse, which has been achieved with fiber compression down to the 2-cycle range [4]. Meanwhile, the self-compression of an energetic 800-nm pulse to sub-cycle length in a noble-gas filled hollow-core photonic crystal fiber has been widely predicted [5,6] and inferred by experimental results. Using this scheme, few-cycle pulses have been demonstrated for some time [7], and very recently single-cycle pulse generation was also demonstrated [8]. With the rising availability and quality of such intense ultrashort pulses, which are expected to be ideal for the generation of attosecond pulses, we aim to investigate theoretically and in depth the electronic responses driven by them.

The radiative response of HHG is well understood through rescattering theory [9]. In each rescattering event, the electrons are ionized by the peak of the external field, and then driven back by the subsequent field, striking their parent ionic cores and emitting high-energy photons in an attosecond pulse. The attosecond burst is created roughly 3/4 cycle after the initial tunnel ionization. The electron trajectory and thus the emitted photon energy can be estimated classically, which yields the cut-off frequency of the radiated spectrum—typically tens to hundreds of eV. More recently, the detailed structure of the target has been incorporated into the rescattering theory successfully, giving new opportunities in enhancement of HHG and in dynamic imaging [10,11,12]. Despite such developments, so far the studies of HHG driven by ultrashort pulses [13,14,15,16,17,18,19] have mostly focused on the broadband harmonics and coherent attosecond emission near the cut-off region, disregarding harmonic emission below or near the ionization-threshold $I_p$, where the electronic dynamics are more strongly distorted by the atomic potential and bound states. On the other hand, a few papers specifically study the below-threshold harmonics typically in the vacuum-ultraviolet (VUV) frequencies, focusing mainly on the phase of lower-order harmonics affected by the atomic potential [20,21] and on the atomic bound-state resonances in harmonic spectra [22,23], albeit driven by longer pulses.

In this report, we simulate the dipole response of a hydrogen atom to an intense sub-cycle near-infrared pulse with an *ab initio* calculation for the single-active-electron time-dependent Schrödinger equation (SAE-TDSE) [24], and examine the resulting radiation spectra. In the short-pulse limit, a delayed, monochromatic, strong radiation signal exists just above $I_p$, which behaves very differently from the typical HHG signal or the Kerr-like perturbative dipole response. We believe this novel effect, which exists only for ultrashort driving pulses in the strong-field regime, to be measurable and general to all atomic species. We additionally perform a calculation with the semi-classical strong field approximation (SFA) [25] and observe the same phenomenon. Since bound electronic states are excluded in the SFA calculation, this surprising resonance-like behavior presents an unprecedented radiative mechanism and renews our understanding of strong-field physics, and could lead to new possibilities in controllable VUV light sources.

The physical system is described as a single electron in a hydrogenic potential and in an arbitrarily linearly-polarized, finite-duration external electric field. The TDSE calculation is supplied by the finite difference SAE-TDSE package, CLTDSE [24]. The implementation of SFA follows established techniques [25]. The single-atom radiation signal is defined by dipole acceleration, i.e., $R(t) = d^2D(t)/dt^2$ where $D(t)$ is the time-dependent dipole moment. The radiation spectrum is the intensity of the Fourier transform of $R(t)$.

## 2. Radiation spectra

With the compression method demonstrated in Ref. [7] and a comprehensive propagation model that has been extensively tested against experiments [26], we estimate that a 2.5 µJ pulse at 800 nm can be compressed from 35 fs to 1.5 fs—less than a single cycle (2.67 fs)—in a 8-cm long, 27-µm core-diameter kagome-type photonic crystal fiber filled with 3 bar of argon. The (cycle-averaged) peak field intensity on the fiber axis at the output end is $3.5 \times 10^{14}$ W/cm$^2$, which is a typical intensity level used to drive HHG. The carrier-envelope phase (CEP) is preserved during the compression. Although there is still some residual field along the 35-fs-long "pedestal" of the compressed pulse, its intensity is at least 4 times lower than at the sub-cycle pulse center, so it has very little impact on the strong-field process.

In the following we discuss the result of the TDSE calculation. Given the above experimental background, for simplicity and generality in this theoretical study, we take the driving pulse to have a Gaussian temporal envelope with FWHM duration of 2 fs and peak envelope intensity $3 \times 10^{14}$ W/cm$^2$ (field amplitude $E_0 = 4.7 \times 10^{10}$ V/m). The ionization potential is 13.6 eV for hydrogen atoms. The driving pulses for CEP values 0 and π/2, together with the calculated radiation spectra, are shown in Fig. 1(a) and (b).

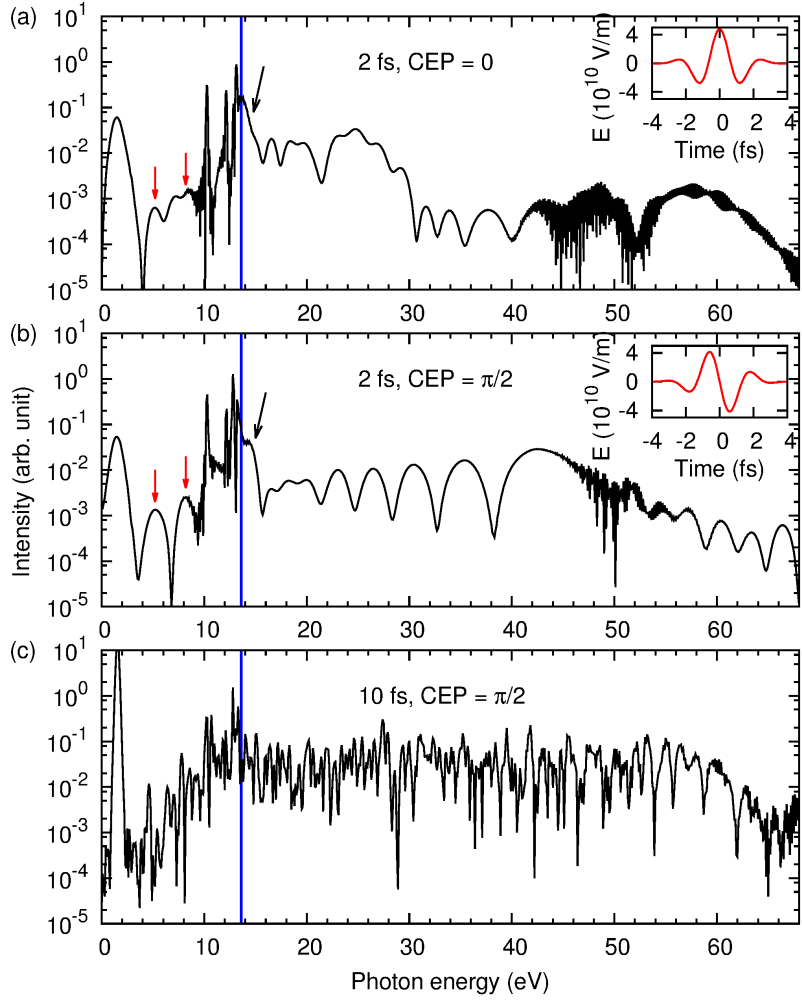

**Figure 1.** Dipole radiation spectra driven by a 2 fs pulse with CEPs of (a) 0 and (b) π/2, and (c) by a 10 fs pulse with CEP = π/2; the insets show the driving electric fields for (a) and (b) respectively. The blue vertical lines indicate the ionization potential at 13.6 eV. The 3rd and 5th harmonics are marked by red arrows. The black arrows point to the appearance of the threshold dipole response (TDR).

The classical picture of HHG driven by a monochromatic field [9] predicts that the cut-off frequency of the radiation spectrum is $I_p + 3.17\, U_p$, where $U_p$ is the ponderomotive energy. However, because the actual ultrashort driving pulse is strongly distorted compared to a monochromatic oscillation, each sub-peak in intensity is responsible for different cut-offs in the spectrum. As seen in Fig. 1, the two CEPs change the driving waveforms totally, producing quite different overall spectral landscapes. Such strong CEP dependence in short-pulse driven HHG has been experimentally demonstrated [27] and theoretically analyzed [17,18,19]. In our particular case, for CEP = 0 there are two cut-offs at 58 eV and 25 eV, corresponding to the intensity peaks at 0 and ±1.2 fs, while for CEP = π/2 the cut-off at 43 eV corresponds to the intensity peaks at ±0.6 fs. The CEP shifts the strongest emitted frequency by nearly 20 eV. This represents dramatic control of the VUV emission by the CEP of the driving field, which is itself set by the CEP of the laser pulse before compression.

In the ultrashort driving field, spectral interference is very weak because the radiation is generated in a single return of electrons. As a result, the spectrum above $I_p$, which is determined by the non-perturbative dipole response (NPDR), is continuous and the harmonic orders disappear, as discussed by previous authors [13,14,15]. On the other hand, the low-order harmonics are mainly created by the perturbative dipole response (PDR), which is Kerr-like and nearly instantaneous. The 3rd harmonic at 5 eV and the 5th harmonic at 8 eV are clearly distinguishable and insensitive to the CEP as seen in Fig1. The intensities and phases of these below-threshold harmonics have been reported in detail [20,21]. Between the two regimes, the spectrum exhibits two other prominent components in the vicinity of $I_p$. First, a group of strong emission lines appears from 10 eV up to $I_p$, representing the atomic resonances. Such resonances in the HHG spectrum have been reported both experimentally and theoretically in recent years [14,15]. Second, there exists a 1.5-eV-wide emission band just above $I_p$, whose signal intensity is roughly an order of magnitude higher than the surrounding spectrum, if resonance peaks are not taken into account. This feature corresponds to the mechanism dubbed hereafter as the threshold dipole response (TDR), which will be explained in details later in this article.

To fully appreciate the importance of pulse length for the spectral characteristics shown by the 2 fs pulse, in Fig. 1(c) we show the radiation spectrum driven by a 10 fs, CEP = $\pi/2$ pulse. With less than 4 optical cycles, the pulse is still considered ultrashort in high-field optics. For such a pulse length, the dramatic CEP control and the flat emission in the plateau demonstrated for the 2 fs pulse disappear. Moreover, the PDR and NPDR regimes merge together, and the emission band just above $I_p$ becomes unrecognizable. Practically there is no longer any distinct visual separation between the PDR and the NPDR. Meanwhile, the atomic resonance lines are barely affected by the prolonged driving field. In other words, unlike the actual atomic resonances, the emission band just above $I_p$ exists uniquely for driving pulses approaching sub-cycle durations. This remarkable spectral feature is reported here for the first time.

It is worth noting that in strong-field ionization driven by mid-infrared laser fields, a similar spectral peak in the photoelectron spectrum was discovered experimentally in the low-energy region [28,29]. The phenomenon was attributed to the Coulomb interaction with the ions when electrons tunnel out [29], or the "soft recollision" process where the returning electrons miss the head-on collision with the cores [30]. Although similar in spectral appearance, the feature in the UV radiation spectrum near $I_p$ in our study exists only for the sub-cycle duration of the driving pulse, as shown in Fig. 1 and analyzed later in this paper, while the low-energy photoelectron feature does not depend on the pulse duration.

While our simulations are based on a single atom, reabsorption of radiation by the medium through which the driving pulse propagates could strongly affect how measurable this feature is. The linear absorption cross-section of a gas peaks just above $I_p$ and gradually decreases with higher frequencies. Taking its maximum value near $I_p$ in hydrogen $\sigma = 6.3\times10^{-22}$ m$^2$ [31], and the thickness $L$ = 1 mm and number density $\rho = 8\times10^{23}$ m$^{-3}$ in a typical HHG gas jet [4], for example, the transmission is exp($-\sigma L\rho$) = 60% through the gas. Reabsorption only slightly weakens the near-$I_p$ emission band for the parameter ranges considered here, allowing us to conclude that the TDR is in principle measurable in the macroscopic case.

## 3. Time-frequency analysis

To understand the feature near $I_p$ in the radiation spectrum, we now study the dipole response in the time domain. Since the same essential feature is shown for both CEP values, in the following we only consider the CEP = $\pi/2$ case. In Fig. 2(a), we plot the spectrogram of the dipole radiation $R(t)$ using a 3 fs gating window (bandwidth = 0.6 eV). The plot is zoomed-in to the emission band above $I_p$ and its neighbor regions. For convenience of visualization and discussion, time is defined relative to the center of the driving pulse.

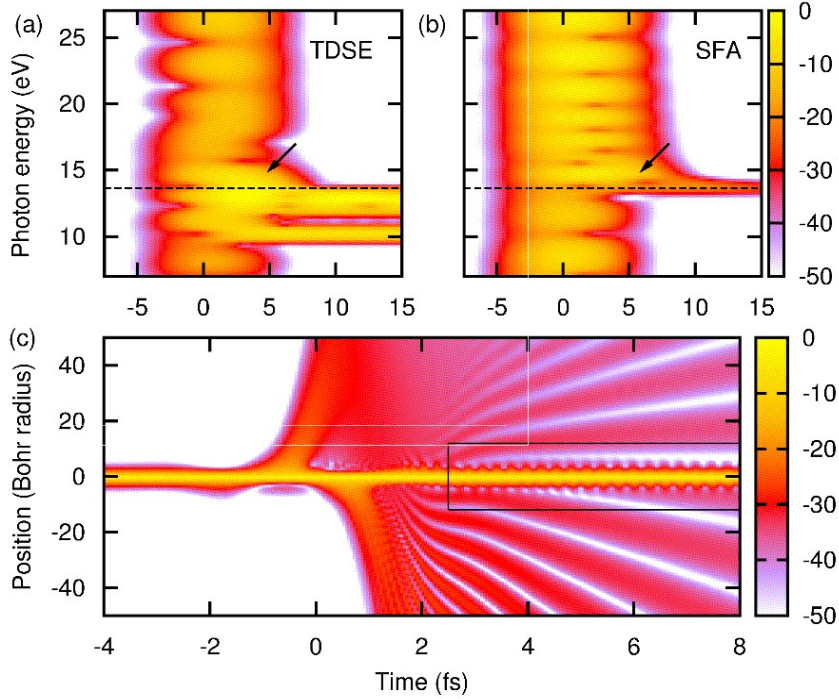

**Figure 2.** Detailed atomic response for the 2 fs, CEP = $\pi/2$ driving pulse used in Fig. 1(b). (a) and (b) are the spectrogram of the dipole radiation $R(t)$, zoomed into the near-$I_p$ region and with a 3 fs gating window, simulated by TDSE and SFA respectively. The ionization threshold is indicated by the horizontal dashed line. The TDR is marked by the black arrow. (c) Probability density of the total wave function in space near the ionic core. The black box highlights the appearance of the TDR. All signals are plotted in dB scale.

In Fig. 2(a), very different characteristics are seen for the traces far below (<10 eV), near (10 to 15 eV), and far above (>15 eV) $I_p$. In the low-frequency region, the PDR dominates and is strongest near the center (time = 0) of the pulse. The response is gradually and slightly delayed in time as the frequency shifts higher. At 8 eV photon energy, the 5th harmonic emission is already delayed from the pulse peak by nearly 1 fs. Far above $I_p$, the NPDR dominates, and the radiation concentrates temporally at about 2 fs, which is the moment when major electron trajectories return to the core [32]. It represents the main rescattering event initiated at -0.6 fs [see the inset of Fig. 1(b)], and its properties are at the focus of HHG studies. Between 10 eV and $I_p$ = 13.6 eV, the horizontal stripes represent the long-lived atomic excited states, which extend outside the plotted time window. Just above $I_p$, a strong radiation signal of about 1.5 eV bandwidth stretches in time from -2 fs to 6 fs (marked by the black arrow in the figure), and is obviously more delayed than both PDR and NPDR. This part of the radiation corresponds to TDR, which has been shown by the emission band just above $I_p$ in the spectrum in Fig. 1. While the NPDR is perceived as the typical HHG response, the TDR behaves very differently, although both of them are non-perturbative by definition.

To further examine the nature of this feature, in Fig. 2(b) we plot the spectrogram of the dipole radiation calculated by SFA. SFA ignores the Coulomb interaction between the ionized electrons and the ionic core, thus precluding the existence of the bound excited states; just below $I_p$ the emission lines of atomic resonances are missing, letting the PDR extend up to $I_p$. However, the typical NPDR and PDR traces (top and bottom of the plot) agree with TDSE very well. Just above $I_p$, the lowest horizontal stripe (indicated by the black arrow in the figure) shifts and stretches out further in time relative to the NPDR trace, which corresponds to the TDR that has been identified in the observation of the TDSE result. The reproduction of the peculiar emission band suggests that the feature does not rely on the presence of the Coulomb potential or atomic excited states, but rather is a general aspect of atomic species. In SFA, an additional very narrow (<1 eV) emission line on top of $I_p$ extends beyond the plotted time window and mimics the excited state resonances in Fig. 2(a). This is due to the nature of the SFA model which can be viewed as an artifact. We will clarify this feature later.

In general, HHG studies in atomic systems should take into account strong field effects such as the Stark effect and the appearance of the Freeman resonance which is caused by the energy levels being shifted by the ponderomotive force. Concerning ultrashort strong fields, the instantaneous ac Stark shifts of the bound states (in attosecond transient absorption) [33,34] and of the continuum (in photoelectron spectroscopy) [35] have recently been investigated. These reports show that the magnitude of strong field effects follow the instantaneous strength of the ultrashort strong field. In our case, the UV radiation is not influenced by these strong field effects because the radiation occurs after the driving field has passed. As clearly seen in Fig. 2(a), the broadband radiation and resonant emission lines all start roughly at 2 fs, when the field strength has already dropped critically [see the electric field plot in the inset of Fig. 1(b)]. Consequently, the resonance energy levels shown in Fig. 1(a) and (b) are not shifted by the driving field and match the measured field-free atomic levels [36].

## 4. Wave function analysis

While the NPDR trace is understood in the classical picture of HHG, the TDR feature has not before been reported or predicted in strong-field studies. In order to identify its origin, we calculated the total wave function based on SFA [25] and show it in Fig. 2(c). We do not plot the TDSE total wave function because the bound state part in space would be very complex around the core, and little information could be extracted. In Fig. 2(c) we plot the evolution of the spatial probability density in time, driven by the same pulse as was used in Fig. 2(a) and (b).

The main torso of the electron cloud, centered at the ionic core (position = 0), distorts slightly downward and upward alternatively at the peaks of the driving field at -1.8 fs, -0.6 fs, 0.6 fs, and 1.8 fs, forming an S-shaped curve in time. Such distortion is instantaneous to the field, and represents the PDR and low-order harmonics. On top of the body of the electron cloud, two streams of electrons are launched, upward at -0.6 fs and downward at 0.6 fs, before they drift in the external field. The stream launched at -0.6 fs is more localized and returns back to the core roughly 2 fs later, generating the main attosecond pulse burst, as shown by the dominant high-photon-energy trace in Fig. 2(a) and (b). The next stream, launched at 0.6 fs, interferes with the previous stream, forming an interference pattern. The acceleration of the returning electrons stops at 2.5 fs when the field abruptly ends, leaving the ionized electrons floating in space. As a result, a spatially broad wave packet—broadband in spectrum as well—is formed at the trailing part of the pulse, which covers the ionic core in space. The slow and near-core electrons in it create a group of small periodic ticks on the upper and lower sides of the ionic core, as seen inside the black-lined box in Fig. 2(c). This periodic structure is associated with the beating signal between the nearly-bound electron states and the ground state when they overlap in space, and the oscillation frequency corresponds to the transition energy between them. This is the mechanism that underlies the TDR.

It should be remembered that the wave function analysis presented above is based on SFA, which ignores the atomic potential felt by the ionized electrons. The ionized electrons that return to the ionic core after the optical pulse has passed become truly "free"', since they are also free from the Coulomb force, and flow with constant velocities near the core. This is represented by the straight stripes after 3 fs in Fig. 2(c), which seem to radiate away from the core with different but constant speeds. Consequently, the slowest electrons will emit radiation with a photon energy of $I_p$ for a very long time while the wave packet gradually thins down. Now let us imagine the wave function in a real atomic system where ionic attraction is present all the time. The major rescattering event is driven by a strong external field, and is only weakly influenced by the atomic potential. It will similarly create a broadband electron wave packet across the ionic core at the trailing edge of the pulse. After the optical field suddenly ends, the electron cloud surrounding the core will find itself in a purely ionic field, meaning that electron-ion collisions take place. Part of the electron cloud is captured by the atomic potential and forms bound states. With a little bit higher kinetic energy, the nearly-bound electrons slowly scatter, i.e., accelerate and decelerate. However, unlike the SFA case, the trajectories of the nearly-bound electrons after the pulse ends are bent by the atomic potential, and their overlap with the ground state is short-lived. Thus although the periodic structure seen in Fig. 2(c) for SFA also exists for a real atomic system, it is disturbed and shortened by the Coulomb force.

With the wave function analysis, the TDR features, shown differently by TDSE and SFA in Figs.2(a) and (b), are better understood. While both methods preserve the main body of the 1.5-eV wide shoulder above $I_p$ very well, for SFA the nearly-bound electrons are totally free after the pulse has passed, and consequently generate a long, monochromatic emission. In TDSE, once the field is over, the system is purely in the atomic potential, and the coherent emission from the nearly-bound electrons is shorter in time, while the emission from bound excited states persists since they are actually stationary states. In other words, while the magnitude and timescale of TDR are different from species to species, the phenomenon is general.

It has been recently demonstrated and discussed that atomic resonances appear as narrow emission lines below $I_p$ [22,23]. In our case, the nearly-bound electrons play the role of the electrons in resonances and give rise to a similar long-lasting, narrowband emission. This is feasible because, within a short timescale, the nearly-bound electrons behave similarly to Rydberg-state electrons, although they eventually will escape to infinity. However, unlike the actual bound state resonances, the appearance of TDR relies on the fact that once the nearly-bound electrons are generated, they are not disturbed by the following external field; this requires the driving pulse to be ultrashort—ideally single- to sub-cycle duration. Most experiments, on the other hand, are conducted with few- to many-cycle driving pulses, so that the TDR feature has not yet been observed.

Interestingly, a very recent experiment revealed that the Kerr response was "resonantly" enhanced by the interband excitation in a dielectric medium [37]. The conduction band therein was not a real excited state but closer to an ionized state. This is analogous to our study where slow electrons right above $I_p$ can mimic the bound electrons temporarily, although in our case this occurs through a rescattering process.

Note that we have not considered spontaneous emission, which would have speeded up the decay rate of the dipole response. Spontaneous emission has, however, little impact over the timescales we are concerned with, and can be independently calculated and introduced into our models if desired. The main observable features and the conclusions of the analysis remain the same.

## 5. Conclusion

With the *ab initio* calculation and the SFA model, the electronic dipole response of an atomic system driven by a sub-cycle pulse in the strong-field regime are simulated and analyzed. In the radiation spectrum, an emission band exists right above the threshold, which disappears when the driving pulse becomes longer. This emission band corresponds to a delayed dipole oscillation starting near the end of, and lasting after, the driving pulse. It is generated by a resonance-like transition between the nearly-bound electrons and the ground state electrons after the field abruptly ends, as suggested by the wave function evolution. The phenomenon relies on an ultrashort driving field of sub-cycle duration, but is general to all gas species. This phenomenon defines a new regime in extreme nonlinear optics, predicted and analyzed in this work for the first time, and could lead to new opportunities in generation and control of VUV pulses.

## Acknowledgements

The authors acknowledge the helpful discussions with Miroslav Kolesik and Jeffrey Brown. WCC thanks the consultation on the CLTDSE code by Cathal Ó Broin, and the feedbacks by Ka Fai Mak, Francesco Tani, and David Novoa.